\documentclass[nofootinbib,superscriptaddress]{revtex4}
\usepackage{amsmath}
\usepackage{amsfonts}
\usepackage{amssymb}
\usepackage{color}
\usepackage{graphicx}
\usepackage{graphics}
\usepackage{adjustbox,lipsum}
\usepackage[paperwidth=210mm,paperheight=297mm,centering,hmargin=2cm,vmargin=2.5cm]{geometry}

\begin{document}

\preprint{OU-HET-978}
\preprint{UT-HET 129}

\title{Gravitational Waves from Phase Transitions in Models with Charged Singlets}

\author{Amine Ahriche}
\email{aahriche@ictp.it}
\affiliation{LMEPA \& Department of Physics, University of Jijel, PB 98 Ouled Aissa, DZ-18000
Jijel, Algeria.}
\affiliation{The Abdus Salam International Centre for Theoretical Physics, Strada
Costiera 11, I-34014, Trieste, Italy.}

\author{Katsuya Hashino}
\email{hashino@het.phys.sci.osaka-u.ac.jp}
\affiliation{Department of Physics, University of Toyama, 3190 Gofuku, Toyama
930-8555, Japan.}
\affiliation{Department of Physics, Osaka University, Toyonaka, Osaka 560-0043,
Japan.}

\author{Shinya Kanemura}
\email{kanemu@het.phys.sci.osaka-u.ac.jp}
\affiliation{Department of Physics, Osaka University, Toyonaka, Osaka 560-0043,
Japan.}

\author{Salah Nasri}
\email{snasri@uaeu.ac.ae}
\affiliation{Department of physics, United Arab Emirates University, Al-Ain, UAE.}
\affiliation{The Abdus Salam International Centre for Theoretical Physics, Strada
Costiera 11, I-34014, Trieste, Italy.}

\begin{abstract}
We investigate the effect of extra singlets on the
electroweak phase transition (EWPT) strength and the spectrum of the
corresponding gravitational waves (GWs). We consider here the standard
model (SM) extended with a singlet scalar with multiplicity N coupled
to the SM Higgs doublet. After imposing all the theoretical and experimental
constraints and defining the region where the EWPT is strongly first
order, we obtain the region in which the GWs spectrum can be reached
by different future experiments such as LISA and DECIGO.\\
\\
\vspace{0.3cm}
\end{abstract}
\maketitle

\section{Introduction}

One of the unsolved puzzles in both particle physics and cosmology
is the existence and the origin of matter antimatter asymmetry in
the Universe; namely why $\eta=(n_{b}-n_{\bar{b}})/n_{\gamma}=5.8-6.6\times10^{-10}$~\cite{Patrignani:2016xqp}. As it
was shown by Sakharov, there are three necessary criteria for generating
an asymmetry between matter and antimatter at high temperature~\cite{Sakharov:1967dj}:
(1) baryon number violation, (2) C and CP violation, and (3) a departure
from thermal equilibrium. A mechanism to generate the asymmetry during the electroweak phase transition is called the electroweak baryogenesis~\cite{Kuzmin:1985mm}. In the standard model (SM) of particle physics, the first two criteria seem to be qualitatively satisfied where baryon number violation occur non-perturbatively via sphaleron process and C and CP symmetries are violated by the weak interaction sector of the SM. However, it has been shown that in the SM the induced CP violation that parametrizes the matter asymmetry is many order of magnitude too small to generate the observed asymmetry. But even with the addition of new sources of CP violation, the third criterion can not be fulfilled with the SM field content as the electroweak phase transition (EWPT) is not strongly first order~\cite{Bochkarev:1987wf}.

The EWPT can get stronger if new bosonic degrees of freedom are invoked around the electroweak scale~\cite{Anderson:1991zb}, 
or higher dimensional effective operators are considered~\cite{Zhang:1992fs}. The EWPT could be strongly first order for 
2HDMs~\cite{Cline:1996mga}, (U)NMSSM~\cite{Apreda:2001us}, and in other models~\cite{Huang:2016cjm}. Whatever the additional 
fields are, a successful electroweak baryogenesis implies that the new physics can be testable at current and future particle
physics experiments. It had been shown that such a deviation of the triple Higgs coupling with respect to its SM value 
is correlated with the EWPT strength~\cite{Kanemura:2004ch}, and therefore, it could be a useful physical observable to probe a strong EWPT at colliders.

Recently, the Advanced Laser Interferometer Gravitational Wave Observatory
(aLIGO)~\cite{Abbott:2016blz} detected gravitational waves (GWs).
Besides confirming the prediction of Einstein's general theory of
relativity, the GWs opened a new exciting window to probe and test
new physics beyond the standard model of particle physics and cosmology.
For instance, stochastic GWs backgrounds can be generated during first
order cosmological phase transitions (PTs)~\cite{Kamionkowski:1993fg}
which can be within the reach of the near future space-based interferometers,
such as LISA~\cite{Seoane:2013qna}, DECIGO~\cite{Kawamura:2011zz},
BBO~\cite{Corbin:2005ny}, in addition to the Chinese projects TAIJI~\cite{Gong:2014mca}
and TianQin~\cite{Luo:2015ght}. The GWs from first order EWPT that are 
detectable by current/future experiments can be used to probe extended models beyond 
the SM~\cite{Kehayias:2009tn}. Expected uncertainties in future space-based interferometers on parameters of the models 
with the strongly first order phase transition can be partially estimated by the Fisher matrix analysis, 
which is essentially based on a Gaussian approximation of the likelihood function~\cite{Hashino:2018wee}.

In this letter, we consider the SM extended by extra massive scalar(s) 
with ad hoc multiplicity $N$, that is (are) coupled to the Higgs
doublet. Then, after imposing different theoretical and experimental
constraints, we will estimate the effect on the EWPT strength as well as on
the GWs properties, and whether they are in the reach of current/future experiments.

In section II, we present the SM extended by a new scalar
with multiplicity $N$, and discuss different experimental
constraints on the model parameters. We show different aspect of the
a strong first order EWPT in section III. A brief description of the
gravitational waves that could be produced during a strong first order
EWPT is discussed in section IV. In section V, we show and discuss our numerical
results. Finally, we conclude in section VI.

\section{The Model: SM extended by singlets}

In our model, the SM is extended by extra scalar fields $S_{i =1,2,..n}$, 
which transform as $(1, Q_S)$ under $SU(2)_L \times U (1)_Y$, and for simplicity we assume they 
have the same mass. Defining $S\sim (S_1, S_2, .. S_n)$, these scalars 
amount for $N = 2 \times n$ degrees of freedom since $S_i$ are complex scalar fields. The Lagrangian reads,

\begin{equation}
{\cal L}={\cal L}_{SM}+\left|D_{\mu}S\right|^{2}-V(H,S),
\end{equation}
with $D_{\mu}S=\partial_{\mu}S-iQ_{S}g'B_{\mu}S$ and $B_{\mu}$ is
the $U(1)_{Y}$ gauge filed. The tree-level scalar potential is given by 
\begin{eqnarray}
V(H,S) & = & -\mu^{2}|H|^{2}+\frac{\lambda}{6}|H|^{4}+\mu_{S}^{2}\left|S\right|^{2}+\frac{\lambda_{S}}{6}\left|S\right|^{4}+\omega|H|^{2}\left|S\right|^{2},\label{VQ}
\end{eqnarray}
with $H^{T}=(\chi^{\pm},(\phi+i\chi^{0})/\sqrt{2})$ is the Higgs doublet
and $\lambda$, $\lambda_{S}$ and $\omega$ are dimensionless scalar
couplings. The renormalized one-loop effective potential at zero temperature
is given a la $\overline{{\rm MS}}$ scheme by~\cite{Martin:2001vx}
\begin{eqnarray}
V_{1-loop}^{T=0}\left(\phi\right) & = & -\frac{\mu^{2}}{2}\phi^{2}+\frac{\lambda}{24}\phi^{4}+\frac{1}{64\pi^{2}}\sum_{i=all}n_{i}m_{i}^{4}\left(\phi\right)\left(log\frac{m_{i}^{2}\left(\phi\right)}{\Lambda^{2}}-c_{i}\right),\label{Veff-1}
\end{eqnarray}
where $n_{i}$ and $m_{i}$ are the field multiplicities and the field-dependent
masses, respectively. The numerical constants $c_{i}$ for scalars and fermions
(gauge bosons) is $3/2$ ($5/6$) and the renormalization scale is
taken to be the measured Higgs mass $\Lambda=m_{0}=125.09~GeV$. The field dependent masses 
can be written in the form $\mu_{i}^{2}+\alpha_{i}\phi^{2}/2$,
i.e., 
\begin{eqnarray}
m_{\phi}^{2}(\phi) & = & -\mu^{2}+\frac{\lambda}{2}\phi^{2},\,m_{\chi}^{2}(\phi)=-\mu^{2}+\frac{\lambda}{6}\phi^{2},\,m_{W}^{2}(\phi)=\frac{1}{4}g^{2}\phi^{2},\nonumber \\
m_{Z}^{2}(\phi) & = & \frac{1}{4}(g^{2}+g'^{2})\phi^{2},
\,m_{S}^{2}(\phi)=\mu_{S}^{2}+\frac{1}{2}\omega \phi^{2},\,m_{t,b}^{2}(\phi)=\frac{1}{2}y_{t,b}^{2}\phi^{2},
\end{eqnarray}
where $\upsilon=246.22$~GeV is the Higgs vacuum expectation value,
$g$, $g'$ and $y_{t,b}$ are the gauge and Yukawa couplings. The parameters 
$\mu^{2}$ and $\lambda$ can be expressed in terms of the Higgs mass and the EW vacuum as 
\begin{eqnarray}
\lambda & = & \frac{3m_{0}^{2}}{\upsilon^{2}}-\frac{3}{32\pi^{2}}\sum_{i}n_{i}\alpha_{i}^{2}\left(log\frac{m_{i}^{2}}{m_{0}^{2}}+c'_{i}\right),\,\mu^{2}=\frac{1}{6}\lambda\upsilon^{2}+\frac{1}{32\pi^{2}}\sum_{i}n_{i}\alpha_{i}m_{i}^{2}\left(log\frac{m_{i}^{2}}{m_{0}^{2}}-c''_{i}\right),\label{eq:Tad}
\end{eqnarray}
where $c'_{i}$ and $c''_{i}$ for scalars and fermions (gauge bosons)
are 0 (1/3) and 1 (1/3), respectively. According
to (\ref{eq:Tad}), the contribution of a heavy scalar makes the Higgs
quartic coupling at tree-level smaller until it gets vanished. This
implies a new constraint on the space parameter as 
\begin{equation}
N\omega^{2}log\frac{m_{S}^{2}}{m_{0}^{2}}<\frac{32\pi^{2}m_{0}^{2}}{\upsilon^{2}}-\sum_{i=SM}n_{i}\alpha_{i}^{2}\left(log\frac{m_{i}^{2}}{m_{0}^{2}}+c'_{i}\right)\sim88.813.\label{eq:Lam}
\end{equation}

In case of the scalar $S$ is electrically charged, i.e. $Q_{S}\neq0$, the Higgs decay width $\phi\rightarrow\gamma\gamma$
could be modified as
\begin{eqnarray}
R_{\gamma\gamma}^{\phi} & = & \frac{\Gamma(\phi\rightarrow\gamma\gamma)}{\Gamma^{SM}(\phi\rightarrow\gamma\gamma)}=\left|1+\frac{NQ_{S}^{2}\omega\upsilon^{2}}{2m_{S}^{2}}\frac{A_{0}\left(4m_{S}^{2}/m_{0}^{2}\right)}{\frac{4}{3}A_{1/2}\left(4m_{t}^{2}/m_{0}^{2}\right)+A_{1}\left(4m_{W}^{2}/m_{0}^{2}\right)}\right|^{2},\label{Rgg}
\end{eqnarray}
where the functions $A_{i}$ are given in~\cite{Chen:2013vi}. This
ratio should be in agreement with the recent combined results $1.02_{-0.12}^{+0.09}$
of ATLAS and CMS~\cite{ATLAS:2018doi}. The existence of charged
scalar(s) could modify the oblique parameters, namely the $\Delta S$ parameter,
however, this contribution is not significant since the singlet does
not couple to the $W^{\pm}$ gauge bosons. We will consider this as
constraint on the model free parameters $\{m_{S},~\omega,~N\}$. The
existence of an extra scalar that couples to the Higgs can modify
its triple coupling with respect to the SM, as~\cite{Ahriche:2013vqa}
\begin{eqnarray}
\Delta_{\phi\phi\phi}=\frac{\lambda_{\phi\phi\phi}-\lambda_{\phi\phi\phi}^{SM}}{\lambda_{\phi\phi\phi}^{SM}}=\frac{N\omega^{3}\upsilon^{2}}{32\pi^{2}m_{S}^{2}}\frac{\upsilon}{\lambda_{\phi\phi\phi}^{SM}}=3\times10^{-6}N\left(\frac{\omega}{0.1}\right)^{3}\left(\frac{m_{S}}{300\,GeV}\right)^{-2},\label{Dhhh}
\end{eqnarray}
with the one-loop triple Higgs coupling in the SM is given by~\cite{Kanemura:2002vm} 
\begin{equation}
\lambda_{\phi\phi\phi}^{SM}\simeq\frac{3m_{0}^{2}}{\upsilon}\left[1-\frac{m_{t}^{4}}{\pi^{2}\upsilon^{2}m_{0}^{2}}\right].
\end{equation}

It is expected that a significant deviation in the triple Higgs coupling
(\ref{Dhhh}) can be tested at the LHC~\cite{CMS:2013xfa} as well
as at future lepton colliders such as the International Linear Collider
(ILC)~\cite{BrauJames:2007aa}, the Compact Linear Collider (CLIC)~\cite{CLIC},
or the Future Circular Collider of electrons and positrons (FCC-ee)~\cite{FCC-ee}.
At the high-luminosity LHC, the triple Higgs coupling can be constrained
in less than 100\%~\cite{LHC-hhh}, while at the ILC it can be measured
with a precision of 10\%~\cite{ILC-hhh}.

Another interesting issue that should be considered, is that if the
singlet scalar is not electrically neutral ($Q_{S}\neq0$), it must be unstable.
Depending on the electric charge $Q_{S}$, the singlet scalar field
S could be coupled to SM charged leptons and neutrinos (and/or to
quarks) as in many neutrino mass models that involve charged scalars.
For instance, if $|Q_{S}|=1$, the term $SL^{T}CL$ is allowed in
the Lagrangian of the model\footnote{If the model contains right handed neutrinos, as it is expected to
be the case in many SM extensions that are motivated by neutrino mass
and dark matter, then a coupling of the form $Se_{R}^{T}CN_{R}$ is
allowed.}, with $C$ the charge conjugation operator, and therefore $S^{\pm}$
will decay very quickly. In the case where $|Q_{S}|=m>1$, one can
add higher dimensional operators of the form $\frac{S(L^{T}CL)^{m}}{\Lambda^{3(m-1)}}$,
with $\Lambda>TeV$ is the scale above which the theory needs UV completion\footnote{If the scalar charge $Q_{S}$ is even, i.e. $Q_{S}=2k$, then the
operator $\frac{S(e_{R}^{T}Ce_{R})^{k}}{\Lambda^{3(k-1)}}$ is allowed}. For $Q_{S}>3$, one expects the life time of the charged scalar
$S$ to be very large unless the UV scale is around TeV.

The charged singlets could be produced at both leptonic and hadronic colliders \'a la Drell-Yan 
($q\bar{q} (e^+e^-) \to \gamma / Z \to S^+ S^-$), and seen as a pair of charged leptons and missing 
energy. By searching for dilepton~\cite{Ahriche:2014xra,Guella:2016dwo} and trilepton~\cite{Cherigui:2016tbm} 
signals would be a very interesting way to probe the effect of these charged scalars. Another approach to constraint 
the charged scalars masses couplings to SM fermions at LHC uses dimension-5 operators since the renormalizable interactions 
of the fields $S^{\pm}$ with the SM leptons are already constrained by the lepton charged current rare decay data~\cite{Patrignani:2016xqp}. 
As a result, a charged singlet charged scalar with mass below 260 GeV will be excluded by the 13 TeV LHC with an integrated luminosity of 
3000 $fb^{-1}$~\cite{Cao:2017ffm}. At future leptonic colliders, the use of dimension-5 operator can lead lead to stronger bound. For 
instance, $m_S < 145~ GeV (170 GeV)$ can be excluded by the ILC-350 (ILC-500) with ${\cal L}=1~ab^{-1}$, whereas CEPC ($\sqrt{s}=250~GeV$ 
and ${\cal L}=5~ab^{-1}$) can be exclude charged singlet scalars lighter than 112 GeV~\cite{Cao:2018ywk}.

\section{Electroweak Phase Transition Strength}

The one-loop effective potential at finite temperature is given by~\cite{Th}
\begin{eqnarray}
V_{eff}^{T}\left(\phi,T\right) & = & V_{1-loop}^{T=0}\left(\phi\right)+T^{4}\sum_{i=all}n_{i}J_{B,F}\left(m_{i}^{2}/T^{2}\right),\label{Veff}\\
J_{B,F}\left(\alpha\right) & = & \frac{1}{2\pi^{2}}\int_{0}^{\infty}r^{2}\log\left[1\mp\exp(-\sqrt{r^{2}+\alpha})\right].\label{JBF}
\end{eqnarray}

A higher order thermal contribution described by the so-called daisy
(or ring) diagrams~\cite{ring}, can be considered by replacing the
scalar and longitudinal gauge field-dependent masses in (\ref{Veff})
by their thermally corrected values~\cite{Gross:1980br}. The thermal
self-energies are given by 
\begin{eqnarray}
\Pi_{\phi} & = & \Pi_{\chi}=\left(\frac{1}{2}\lambda+\frac{3}{16}g^{2}+\frac{1}{16}g^{\prime2}+\frac{1}{4}y_{t}^{2}+\frac{N}{12}\omega\right)T^{2},\,\Pi_{W}=\frac{11}{6}g^{2}T^{2},\nonumber \\
\Pi_{B} & = & \left(\frac{11}{6}+\frac{N}{3}Q_{S}^{2}\right)g^{\prime2}T^{2},\,\Pi_{W,B}\simeq0,\,\Pi_{S}=\left(\frac{Q_{S}^{2}}{4}g^{\prime2}+\frac{1}{6}\omega+\frac{N+1}{36}\lambda_{S}\right)T^{2}.
\end{eqnarray}

In case of heavy scalar $S$, its contribution to the thermal mass
will be Boltzmann suppressed, then, one can put $N\rightarrow0$.
The transition from the wrong vacuum ($\langle \phi \rangle =0$)
to the true one ($\langle \phi \rangle \neq0$) occurs just
below the critical temperature, $T_{c}$, a temperature value at which
the two minima are degenerate. In the case of a strong first order
EWPT, a barrier exists between the two vacua and the transition occurs
through bubbles nucleation in random points in the space. When the
bubble wall passes through a region at which a net baryon asymmetry
is generated, the $(B+L)$ violating processes should be suppressed
inside the bubble (true vacuum $\langle \phi \rangle \neq0$)
in order to maintain the net generated baryon asymmetry. The criterion
to maintain this generated net asymmetry is given in the literature
by~\cite{Bochkarev:1987wf} 
\begin{eqnarray}
\phi_{c}/T_{c} & > & 1,
\end{eqnarray}
where $\phi_c$ is the Higgs vev at the critical temperature.

Due to the fact that the effective potential at the wrong vacuum $V_{eff}(\phi=0,T)$
does not vanish and is $T$-dependant, then we will take the normalized
effective potential as $V_{eff}(\phi,T)-V_{eff}(\phi=0,T)$ during our analysis.
Indeed, it had been shown that when the value of the thermal effective
potential at the symmetric phase becomes $T$-dependent, the EWPT
dynamics is strongly affected~\cite{Ahriche:2013zwa}.

\begin{figure}[t]
\includegraphics[width=0.5\textwidth]{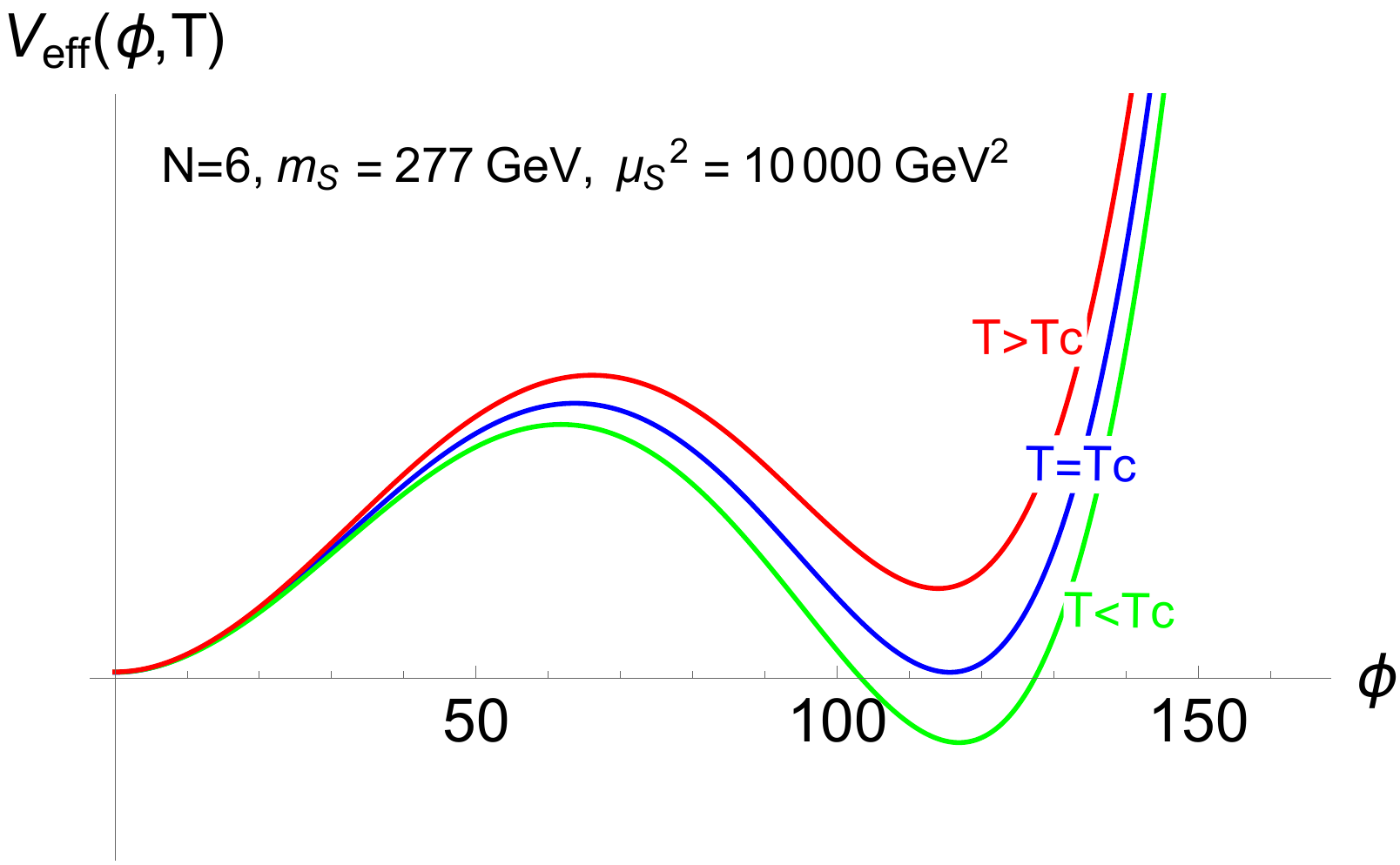}~\includegraphics[width=0.5\textwidth]{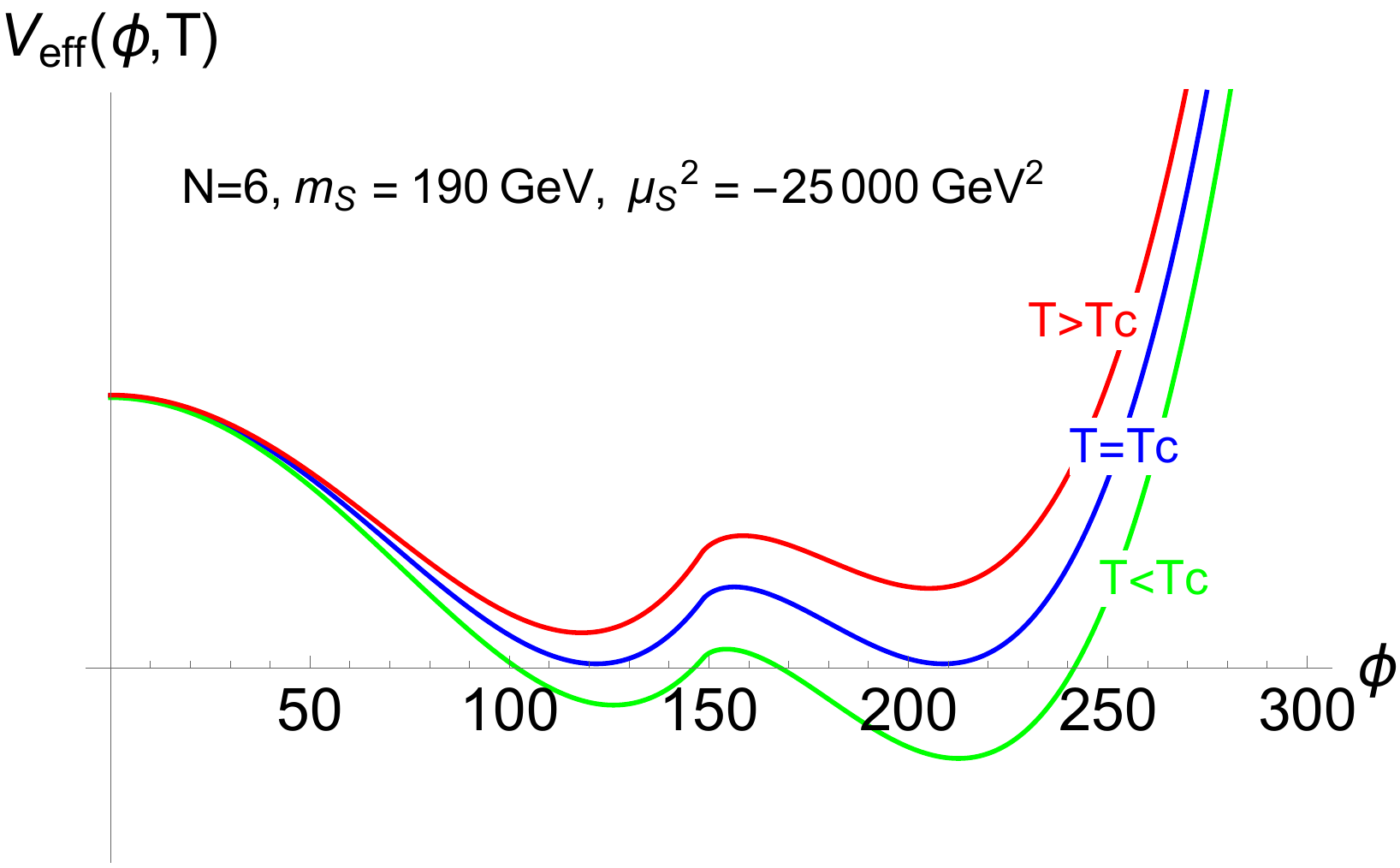}
\caption{The effective potential at different temperature values for type-I PT
(left) and type-II PT (right). Here, we considered the values $Q_{S}=0$ and $\lambda_{S}=0$.}
\label{PT} 
\end{figure}

One has to mention that in the case of second order PT or a crossover,
the true minimum $\langle \phi_{1}\rangle \neq0$ may become
a local minimum below such temperature value. Another new minimum
$\langle \phi_{2}\rangle >\langle \phi_{1}\rangle $
becomes deeper and match the EW vacuum at zero temperature, i.e.,
$\langle \phi_{2}\rangle =246.22\,GeV$ at $T=0$. In such
situation, the transition from $\langle \phi_{1}\rangle $
to $\langle \phi_{2}\rangle $ could only occur via bubbles
nucleation due to the existing barrier between the two minima. Even if the
EW symmetry is already broken (when $\langle \phi_{1}\rangle \neq0$),
the transition from $\langle \phi_{1}\rangle $ to $\langle \phi_{2}\rangle $
may result detectable GWs $\langle \phi\rangle =0 \rightarrow \langle \phi\rangle =\langle \phi_{1}\rangle \neq0$, where the EW symmetry gets broken, 
So we label this transition by type-II PT, while the usual strong first order EWPT by type-I PT. To illustrate
these two pictures, we show in Fig.~\ref{PT} the effective potential
at different temperature values for both type-I PT (left) and type-II
PT (right), respectively.

\section{Gravitational Waves from Phase Transitions}

We analyze the gravitational waves (GWs) spectrum from first order
EWPT in the model. In order to analyze the spectra, we introduce
parameters, $\alpha$ and $\beta$, which characterize the GWs from
the dynamics of vacuum bubble~\cite{Grojean:2006bp}. The parameter
$\beta$ which describes approximately the inverse of time duration of the PT is defined as 
\begin{align}
\beta=-\left.\frac{dS_{E}}{dt}\right|_{t=t_{t}}\simeq\left.\frac{1}{\Gamma}\frac{d\Gamma}{dt}\right|_{t=t_{t}}.
\end{align}
where $S_{E}$ and $\Gamma$ are the Euclidean action of a critical
bubble and vacuum bubble nucleation rate per unit volume and unit
time at the time of the PT $t_{t}$, respectively. We use normalized parameter $\beta$ by
Hubble parameter $H_{T}$ in the following 
analysis: 
\begin{align}
\tilde{\beta}=\frac{\beta}{H_{T}}=T_{t}\frac{d}{dT}\left(\frac{S_{3}(T)}{T}\right)\Bigg|_{T=T_{t}},
\end{align}
where $T_{t}$ is the transition temperature that is introduced by $\Gamma/H_{T}^{4}|_{T=T_{t}}=1$. 
The parameter $\alpha$ is the ratio of the released energy density 
\begin{align}
\epsilon(T)=-V_{eff}(\phi_{t}(T),T)+T\frac{\partial V_{eff}(\phi_{t}(T),T)}{\partial T},
\end{align}
where $\phi_{t}(T)$ is the true minimum at the temperature $T$,
to the radiation energy density $\rho_{rad}=(\pi^{2}/30)g_{\ast}T^{4}$, where 
$g_{\ast}$ is relativistic degrees of freedom in the thermal plasma, at the transition temperature $T_t$:
\begin{align}
\alpha=\frac{\epsilon(T_{t})}{\rho_{{\rm rad}}(T_{t})}.
\end{align}

The GWs production during a strong first order PT occurs via three
co-existing mechanisms: (1) collisions of bubble walls and shocks
in the plasma, where the so-called " envelope approximation" ~\cite{Kosowsky:1992rz}
can be used to describe this phenomenon and estimate the contribution
of the scalar field to the GWs spectrum. (2) After the bubbles have
collided and before expansion has dissipated the kinetic energy in
the plasma, the sound waves could result significant contribution
to the GWs spectrum~\cite{Huber:2008hg}. (3) Magnetohydrodynamic
(MHD) turbulence in the plasma that is formed after the bubbles collision
may also have give rise to the GWs spectrum~\cite{Caprini:2006jb}.
Therefore, the stochastic GWs background could be approximated as
\begin{align}
\Omega_{{\rm GW}}h^{2}=\Omega_{{\rm \phi}}h^{2}+\Omega_{{\rm sw}}h^{2}+\Omega_{{\rm tur}}h^{2}.
\end{align}
The importance of each contribution depends on the PT dynamics, and
especially on the bubble wall velocity. Here, we focus on the contribution
to GWs from the compression waves in the plasma (sound waves) which
is the strongest GWs spectrum among the source of the total GW. A
fitting function to the numerical simulations is obtained as~\cite{Caprini:2015zlo}
\begin{align}
\Omega_{{\rm sw}}(f)h^{2}=\widetilde{\Omega}_{{\rm sw}}h^{2}\times(f/\tilde{f}_{{\rm sw}})^{3}\left(\frac{7}{4+3(f/\tilde{f}_{{\rm sw}})^{2}}\right)^{7/2},
\end{align}
where the peak energy density is 
\begin{align}
\widetilde{\Omega}_{{\rm sw}}h^{2}\simeq2.65\times10^{-6}v_{b}\tilde{\beta}^{-1}\left(\frac{\kappa\alpha}{1+\alpha}\right)^{2}\left(\frac{100}{g_{\ast}^{t}}\right)^{1/3},
\end{align}
at the peak frequency 
\begin{align}
\tilde{f}_{{\rm sw}}\simeq 1.9\times10^{-5}\mbox{Hz}\frac{1}{v_{b}}\tilde{\beta}\left(\frac{T_{t}}{100{\rm GeV}}\right)\left(\frac{g_{\ast}^{t}}{100}\right)^{1/6},
\end{align}
where $v_{b}$ is the velocity of the bubble wall and $\kappa$ is
the fraction of vacuum energy that is converted into bulk motion of
the fluid. Contrary to what it was believed, a succesful electroweak baryogenesis and a sizable gravitational wave signal 
can be achived in the same setup~\cite{Caprini:2007xq}, where the GWs signal is in the reach of BBO and LISA for very 
fine-tuned scenario~\cite{No:2011fi}. Here, since we are interested in probing any possible GWs signal in a general SM 
extension that can map many models, we adopt the value $v_{b}=0.95$. Indeed, in a concreate model, one has to deal with 
many issues like CP violation source(s), the exact bubble wall and plasma velocities.

\section{Numerical Results}

\begin{figure}[h!]
\includegraphics[width=0.5\textwidth]{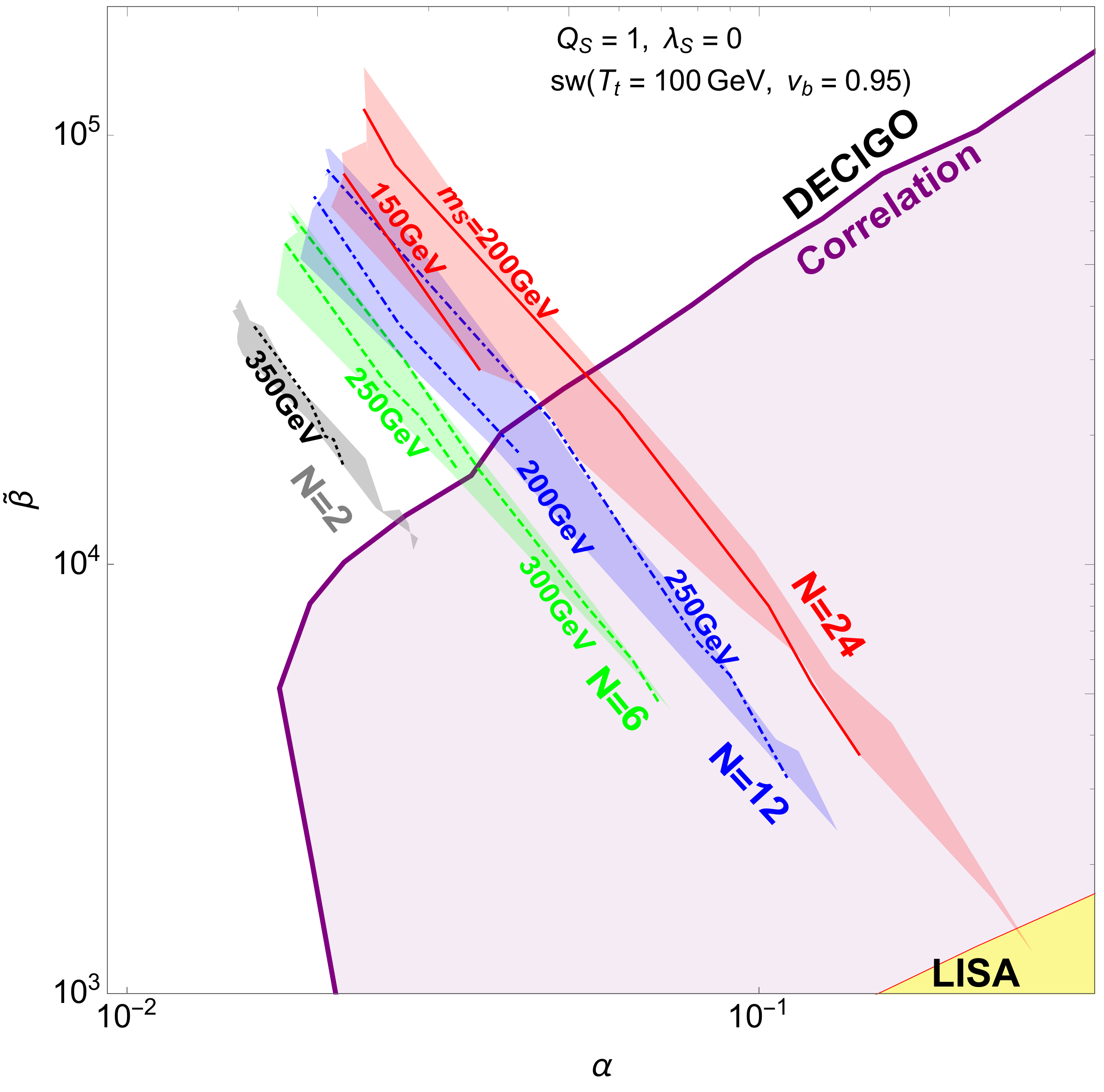}
\caption{The predicted values of $\alpha$ and $\tilde{\beta}$ for different values of 
$\{N,m_{S},\omega\}$ with $Q_{S}=1$, $\lambda_{S}=0$ and $\mu_{S}^{2}\ge0$.}
\label{GWs1} 
\end{figure}

In our numerical results, we have three free parameters $\{N,m_{S},\omega\}$
where we consider the multiplicity values $N=2,6,12,24$, the scalar-Higgs
doublet coupling range $\left|\omega\right|\leq5$, and the scalar
mass to lie in the range $m_{S}\in[100\,GeV,\,550\,GeV]$. In Fig.~\ref{GWs1},
we present the predicted values of $\alpha$ and $\tilde{\beta}$
for different values of $\{N,m_{S},\omega=2(m_{S}^{2}-\mu_{S}^{2})/\upsilon^{2}\}$
with $Q_{S}=1$ and $\lambda_{S}=0$.

In Fig.~\ref{GWs1}, the gray, red, blue and green regions represent $\{N=2,~6,~12,~24\}$,
respectively. We also show the expected sensitivities
of LISA and DECIGO detector configurations are set by using the sound
wave contribution for $T_{t}=100~GeV$ and the bubble wall velocity
$v_{b}$ = 0.95. The colored regions are experimental
sensitivities reached at future space-based interferometers, LISA~\cite{Klein:2015hvg,Caprini:2015zlo,PetiteauDataSheet}
and DECIGO~\cite{Kawamura:2011zz}.

\begin{figure}[h!]
\includegraphics[width=0.5\textwidth]{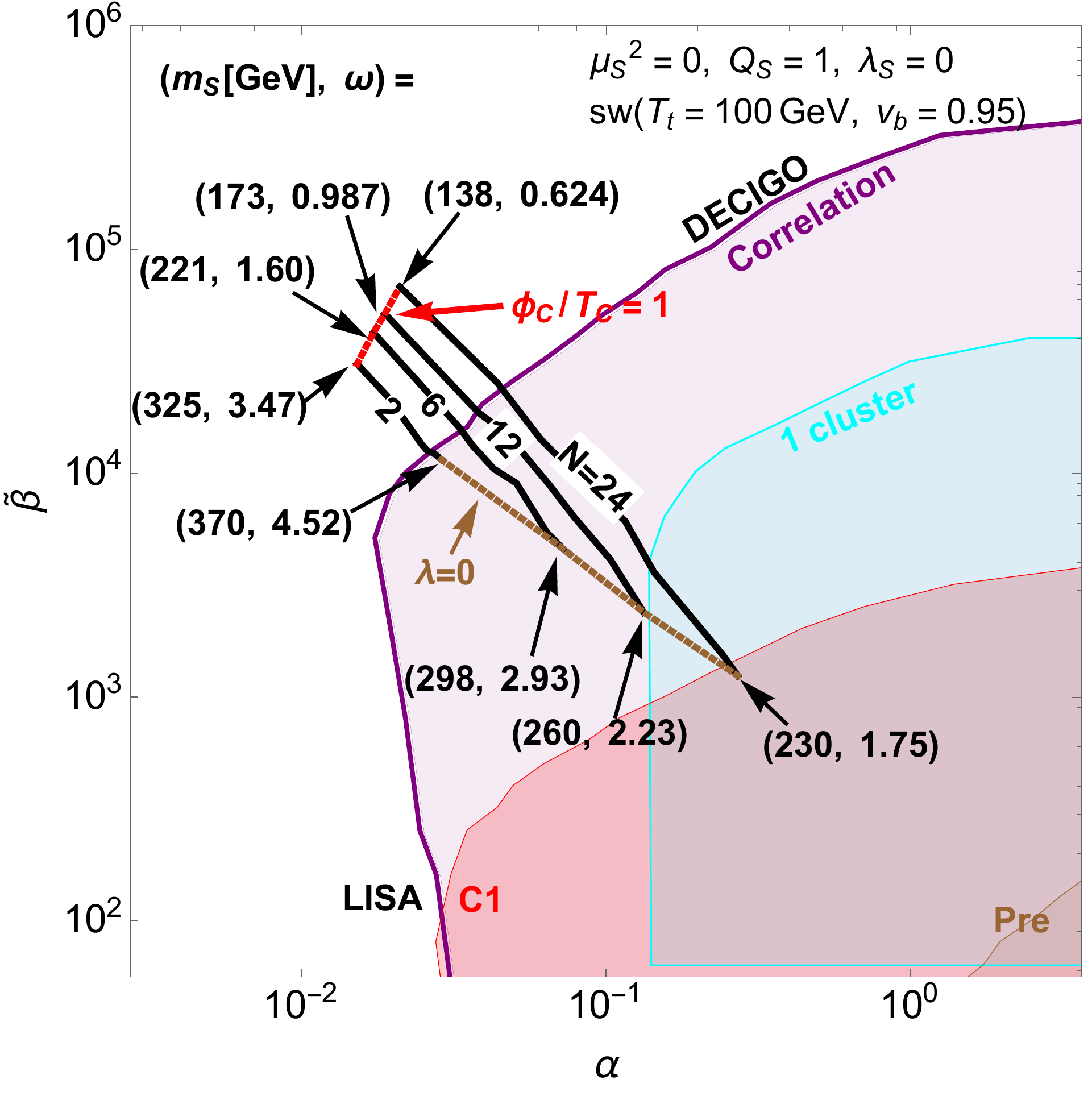}~\includegraphics[width=0.5\textwidth]{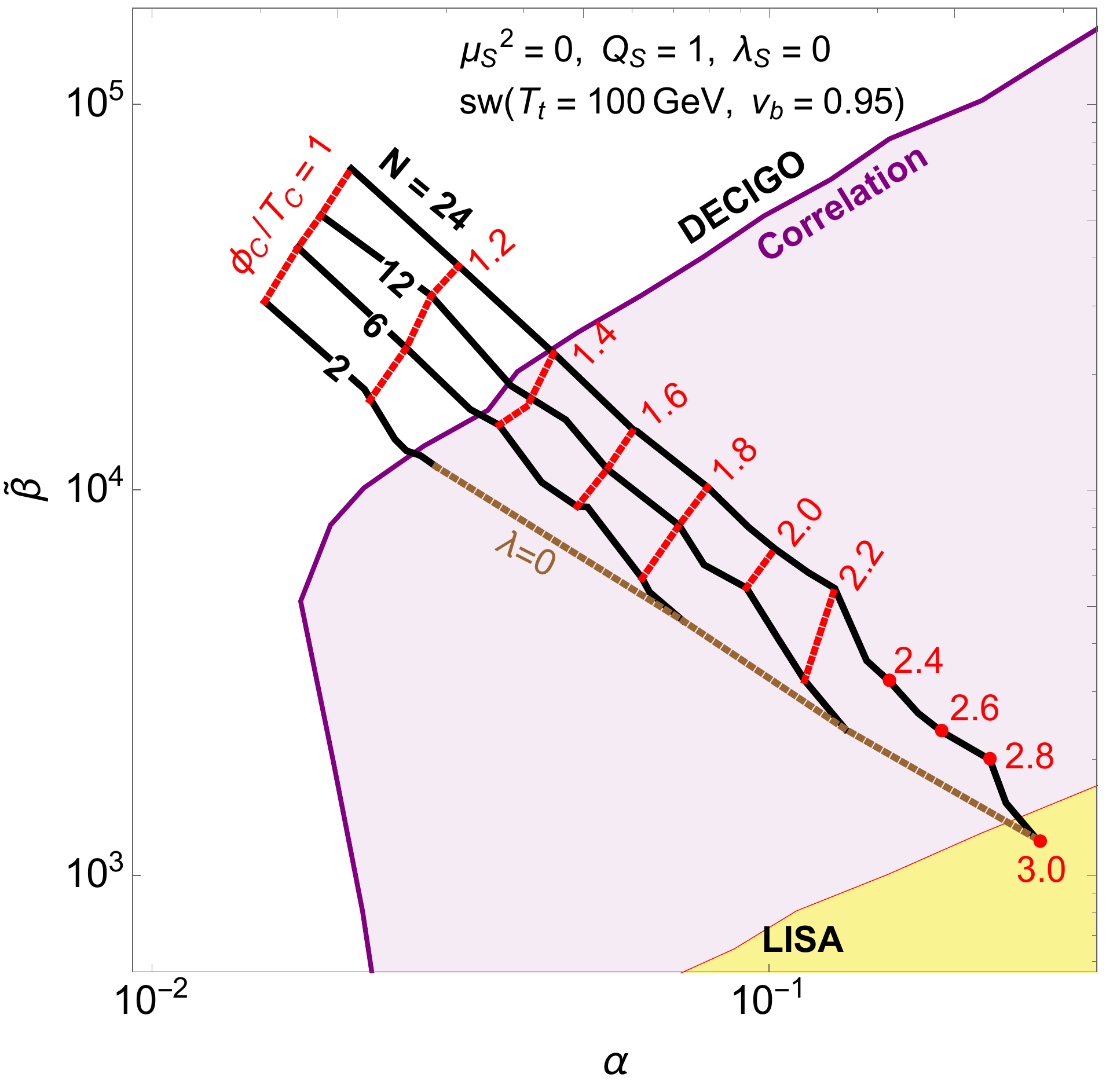}
\caption{The predicted values of $\alpha$ and $\tilde{\beta}$ for different values of 
$\{N,m_{S},\omega\}$ with $Q_{S}=1$, $\lambda_{S}=0$ and $\mu_{S}^{2}=0$.}
\label{GWs2} 
\end{figure}

In Fig.~\ref{GWs2}, we present the special case $\mu_{S}^{2}=0$. Both left and right panels represent the 
same results, where in the left panel we present information about parameter values and constraints, and in the right 
one we show the value of the EWPT strength parameter $\phi_c/T_c$.

In Fig.~\ref{GWs2}, the black curves correspond to the special
case $\mu_{S}^{2}=0$, for $\{N=2,~6,~12,~24\}$, and the upper bound
(in red) on $\tilde{\beta}$ is set by the condition $\phi_{c}/T_{c}=1$.
However, the lower bound (in brown) on $\tilde{\beta}$ is dictated
by the vacuum stability condition (\ref{eq:Lam}). The label "C1" in Fig.~\ref{GWs2}-left, 
corresponds to the configuration of LISA provided in Table.~1
in~\cite{Caprini:2015zlo}, whereas the labels "Pre-DECIGO", "FP-DECIGO" and "Correlation" 
are DECIGO designs~\cite{Kawamura:2011zz}. From this figure, one notices that for stronger PT with $\phi_c/T_c\ge 1.38$ 
the generated GWs can be seen by DECIGO, and for $\phi_c/T_c\ge 2.95$ the corresponding GWs can be seen by LISA.

\begin{figure}[h!]
\begin{centering}
\includegraphics[width=0.5\textwidth]{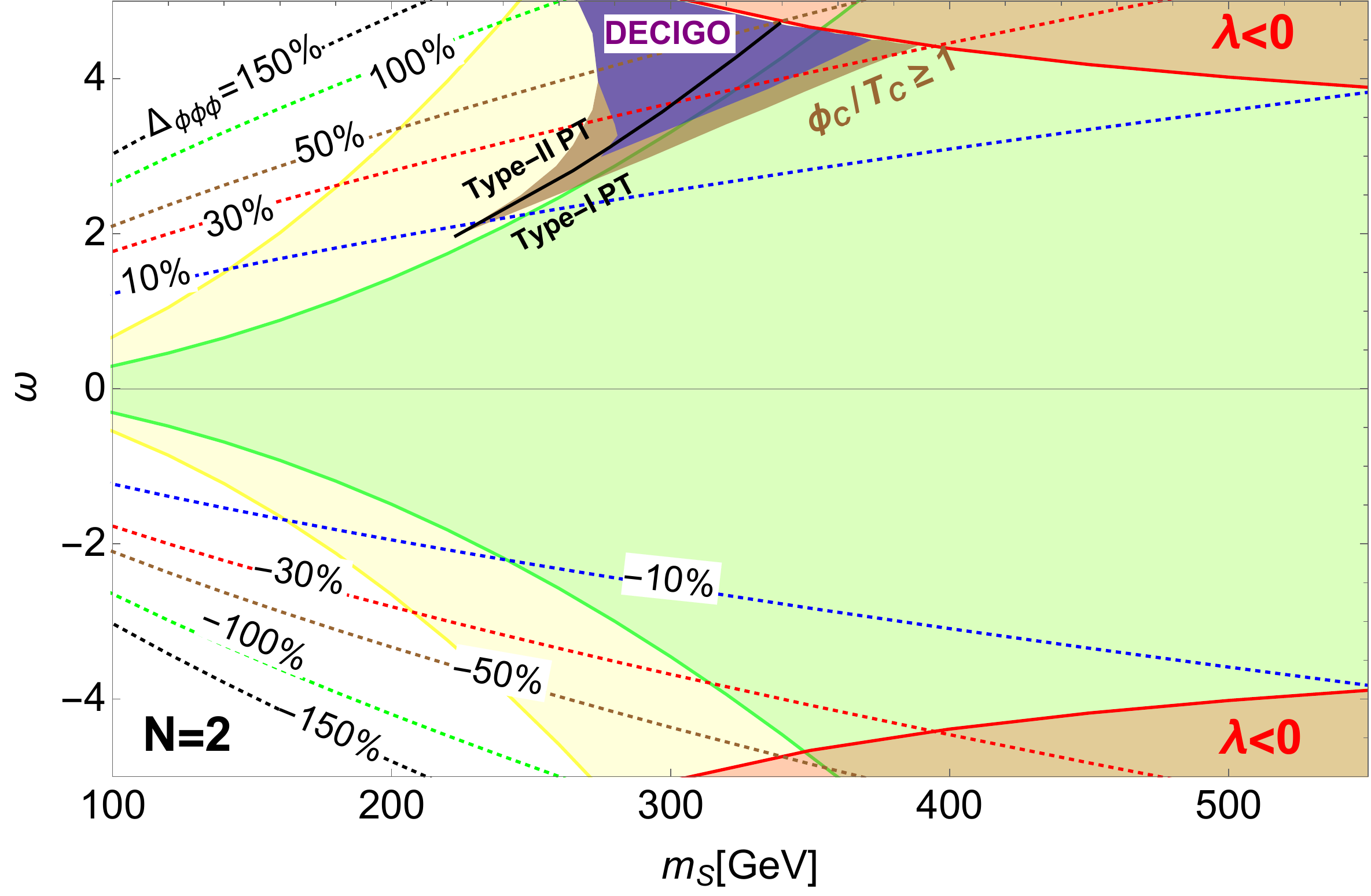}~\includegraphics[width=0.5\textwidth]{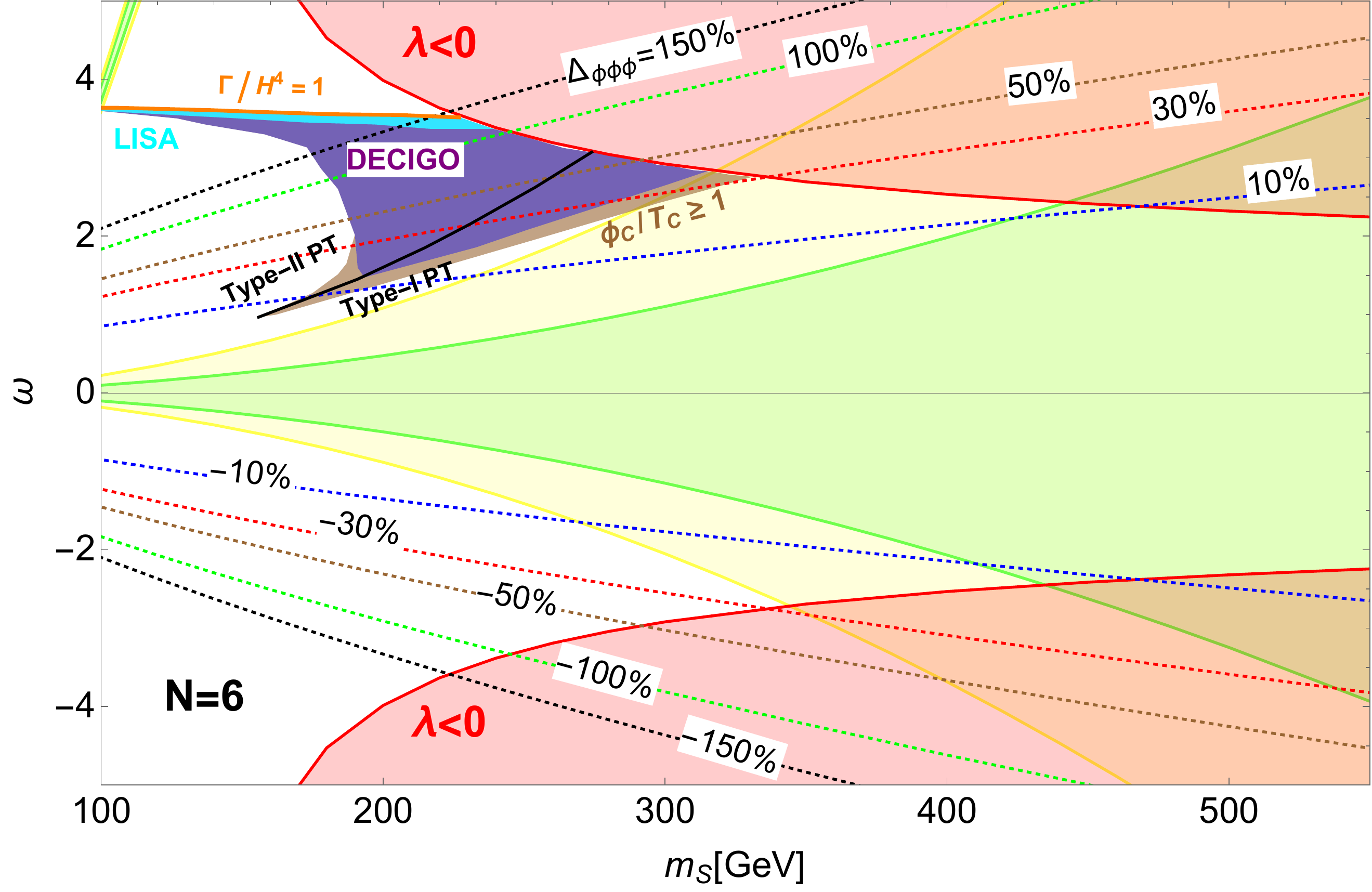}\\
 \includegraphics[width=0.5\textwidth]{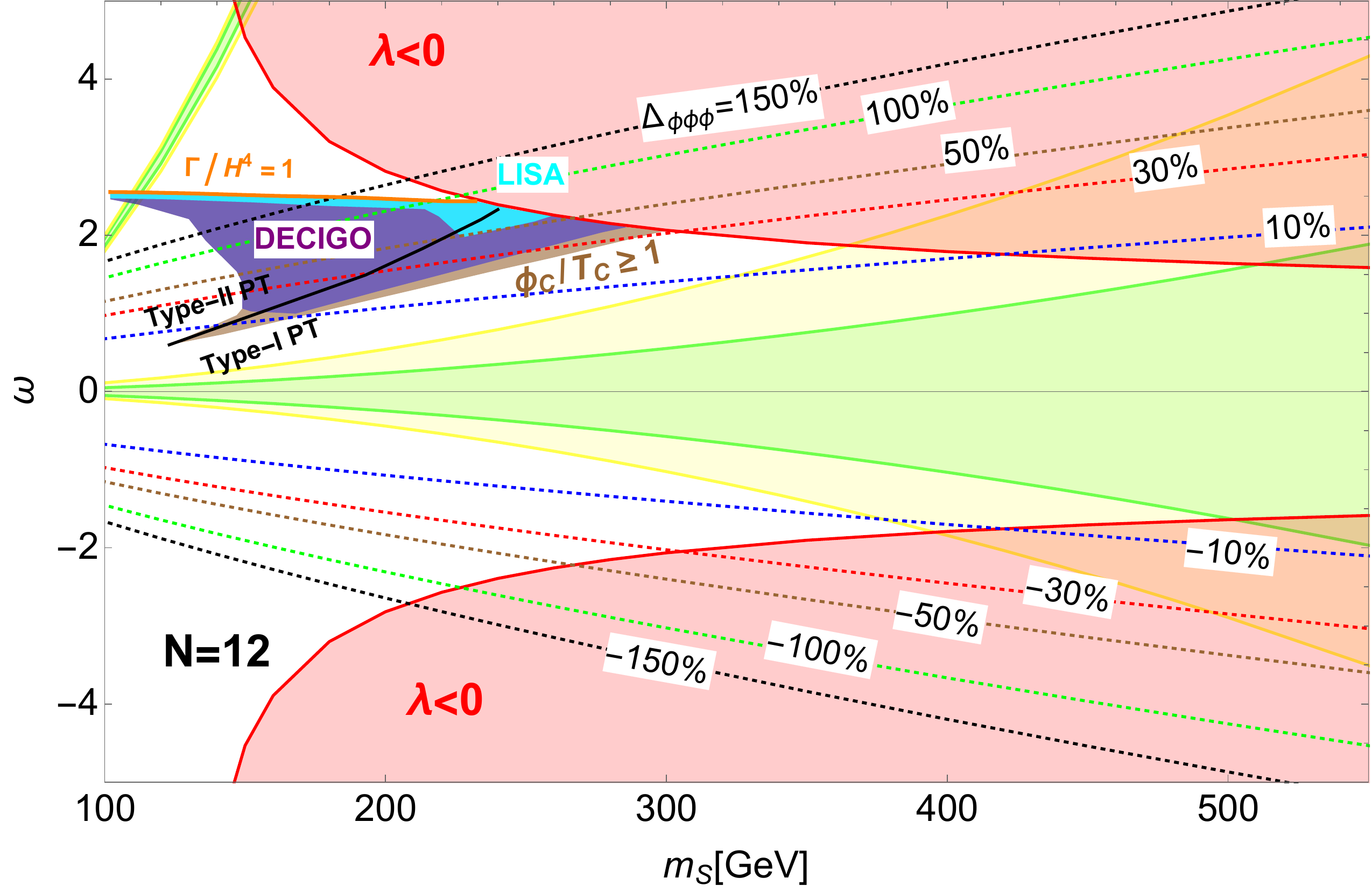}~\includegraphics[width=0.5\textwidth]{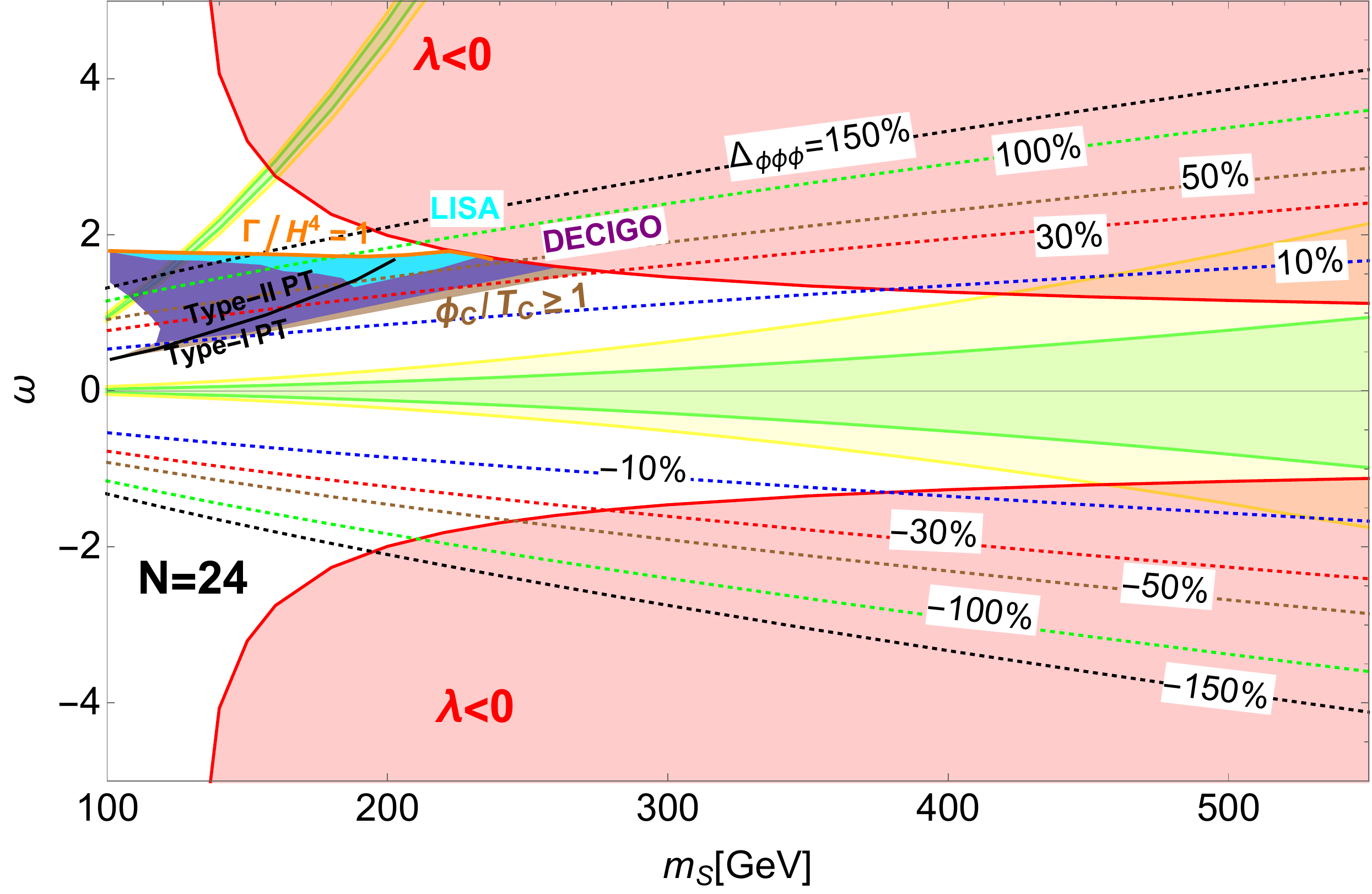} 
\par\end{centering}
\caption{The area of the occurrence of detectable gravitational wave spectrum,
the vacuum stability condition of $\lambda\protect\geq0$ and the
constraint on $R_{\gamma\gamma}^{\phi}$. The dotted lines are the deviation
of the triple coupling (\ref{Dhhh}), and the red regions describe
the vacuum stability condition (\ref{eq:Lam}). The green and yellow
regions represent the current experimental data of $\phi\rightarrow\gamma\gamma$
at the accuracy of $1\sigma$ and $2\sigma$, respectively~\cite{ATLAS:2018doi}. The brown region represents the area of a 
strong PT without possible GWs detectabily, and the black line shows the border of Type-I and Type-II PT.}
\label{omega} 
\end{figure}

In Fig.~\ref{omega}, we show the regions in the plane $\{m_{S},\omega\}$ where there exist a type-I and type-II 
PT for different multiplicity values $N$, together with the constraints mentioned in section II, such as the 
unitarity bound, the vacuum stability condition of $\lambda\ge0$, and $R_{\gamma\gamma}^{\phi}$. In addition, 
we show few contours that express the relative enhancement in the triple Higgs coupling (\ref{Dhhh}). 
In Fig.~\ref{omega}, the green (yellow) regions corresponds to the allowed regions by the experiemental 
measurments of the di-photon Higgs decay for $|Q_S|=1$ at the accuracy of $1\sigma$ ($2\sigma$). The thin ribbon in top-left 
of the cases $N=6,12,24$ is also part of the allowed region by the di-photon Higgs decay, which corresponds to the case 
where the $A_0(m_S^2/4m_0^2)$ term in (\ref{Rgg}) has a value around -2.

From Fig.~\ref{omega}, the GWs detectable region is different from
one multiplicity value to another. If the new singlet is electrically
charged $Q_{S}\ne 0$, the EWPT can not be strongly first order while
fulfilling the constraint from $\phi\rightarrow\gamma\gamma$~\cite{ATLAS:2018doi}
for large multiplicity values $N>2$. In other models that contain scalars with opposite 
couplings to the Higgs doublet this conflict can be evaded, while the EWPT is still strongly 
first order. According to the multiplicity $N$, the scalar mass should lie between 120 GeV and
380 GeV in order to have a strong first order EWPT, i.e., type-I PT, and detectable GWs signal at DECIGO and may be LISA.
However, detectable GWs signal can be observed even for $m_{S}<100~GeV$ for type-II PT, i.e., the EWPT is a second order 
transition (or a cross over); and the baryon asymsetry should be achieved via another mechanism rather than the EWB scenario.

In addition, one has to mention that detectable GWs signal implies
positive enhancement on the triple Higgs coupling (\ref{Dhhh}) with
ratio between 10\% to more than 150\%, depending on the multiplicity
$N$. The main idea one can learn from the results in Fig.~\ref{omega}
is that if the EWPT is strongly first order (type-I PT), the GWs signal would be able to be 
detected by the future space based GW interferometers, in addition to a non-negligible
enhancement in triple Higgs coupling (\ref{Dhhh}). If the EWPT is
crossover or second order, there could exist another transition at
a temperature value smaller that the EWPT where the system vacuum
moves to a new, deeper and larger minimum via bubbles nucleation where
it leaves also detectable signal (type-II PT) the future space based GW interferometers.

\section{Conclusion}

We have focused on the model with additional $N$ isospin singlet scalar fields $S$ which have hypercharge $Q_{S}$. 
In this model, we distinguish two types of EWPT: (1) type-I: the EW symmetry gets broken in one step where the system 
moves from the wrong vacuum ($ \langle\phi \rangle=0$) to the true one ($\langle \phi \rangle\ne 0$); and (2) type-II: 
the EWPT occurs in two steps where the first one is a second order PT (or a crossover), and a new minimum 
($\langle\phi_2\rangle \ne 0$) rather than the existing one ($\langle \phi_2\rangle > \langle\phi_1\rangle$) occurs which 
gets deeper as the temperature gets lower until it becomes the absolute minimum. Then, in the second step, the transition from $\langle\phi_1\rangle$ 
to $\langle\phi_2\rangle$ occurs via bubbles nucleation due to the existing barrier between the two minima.

We noticed also that GWs are large enough to be detected by future space-based interferometers, such as LISA and DECIGO, 
for the condition of strongly phase transition, in either type-I for a successful scenario of electroweak baryogenesis or type-II where baryogenesis 
can be fulfilled via another mechanism, there appear significant deviations in the Higgs couplings 
$\phi\phi\phi$ and $\phi\gamma\gamma$ from the SM predictions. We have discussed how the model can be tested by the synergy between 
collider experiments and GW observation experiments. Consequently, by the detection of the GWs and the measurement
of $\phi\gamma\gamma$, we can test the model. Current experimental data for $\phi\gamma\gamma$ and the parameter region 
where detectable GW occur from the type-I PT can give a constraint on the number of additional singlet scalar $N\leq 6$.

\textbf{Acknowledgments}: K.~H. was supported by the Sasakawa Scientific Research Grant from
The Japan Science Society. S.~K. was supported in part by Grant-in-Aid for Scientific Research on Innovative Areas, 
the Ministry of Education, Culture, Sports, Science and Technology, No.~16H06492 and No.~18H04587, Grant H2020-MSCA-RISE-2014 
No.~645722 (Non-Minimal Higgs).

\end{document}